# Training and pattern recognition by an opto-magnetic neural network

A. Chakravarty[1, 2], J.H. Mentink[1]*, S. Semin[1], A.V. Kimel[1] and Th. Rasing[1]
[1]Radboud University, Institute for Molecules and Materials, Nijmegen, the Netherlands
[2]Charles University, Faculty of Mathematics and Physics, Prague, the Czech Republic
*Correspondence to: j.mentink@science.ru.nl

Neuromorphic computing aims to mimic the architecture of the human brain to carry out computational tasks that are challenging and much more energy consuming for standard hardware. Despite progress in several fields of physics and engineering, the realization of artificial neural networks which combine high operating speeds with fast and low-energy adaptability remains a challenge. Here we demonstrate an opto-magnetic neural network capable of learning and classification of digitized 3x3 characters exploiting local storage in the magnetic material. Using picosecond laser pulses, we find that micrometer sized synapses absorb well below 100 picojoule per synapse per laser pulse, with favorable scaling to smaller spatial dimensions. We thus succeeded in combining the speed and low-dissipation of optical networks with the low-energy adaptability and non-volatility of magnetism, providing a promising approach to fast and energy-efficient neuromorphic computing.

Modern society is strongly data-driven and the rapid growth of digital data use and storage pushes the energy cost consumed by the data centers worldwide[1]. With the emerging of new technologies, such as self-driving cars, AI, and Internet of Things, data generation and use will be further enhanced, increasing the energy costs to unsustainable levels. The vast majority of data-driven applications relies on computations with machine learning methods, which were originally developed to understand the functioning of the human brain. Doing such computations with standard hardware causes massive data transfer between memory and processor, which fundamentally stems from the von Neumann architecture of the computing technology today. One strategy to bypass this bottleneck is neuromorphic computing, which has been developed in several fields of physics and engineering[2-10]. Neuromorphic computing implements brain-inspired computing concepts directly in hardware and processes data in parallel, assisted by local memory, avoiding the energy costly data transfer and thereby drastically reducing the energy consumption. However, despite several important developments in neuromorphic computing[11-13], combining high-speed operation with fast and low-energy training comprises a major challenge.

All-optical artificial neural networks are an appealing approach for high speed and massively parallel data processing, in which data rates are fundamentally limited only by the speed of light. However, adapting the network parameters of optical networks, i.e. adjustment of the synaptic weights, requires significant external data storage[14], external circuitry[15] or even proceeds completely externally[11]. This limits the energy-benefits of all-optical networks, in particular for deep learning applications such as the complex board game Go[16], natural language processing[17] and three-dimensional protein folding[18], where training comprises the majority of the computational effort.

To reduce energy costs and training time while maintaining the processing rates, it is desirable to include local and optically adaptable storage elements within the optical circuitry. Magnetism is already the dominant data storage technology today, where bits of information are stored within the direction of magnetic moments in a nonvolatile and rewritable way. Moreover, in recent years, tremendous progress has been made in the control of magnetism with ultrashort laser pulses, which disclosed switching scenarios that are both faster and more energy-efficient than feasible with external magnetic fields[19-21]. Moreover, in technologically relevant materials such as Co/Pt, multi-shot switching dynamics has been demonstrated[22-25] and exploited to realize continuously adaptable opto-magnetic synaptic weights[26]. Furthermore, it was shown that supervised learning of a network comprising opto-magnetic synapses is feasible using global feedback only, significantly reducing the external data storage. Therefore, combining ultrafast optics with the ultrafast and low-energy optical control of magnetism seems a viable path towards the realization of neuromorphic computing which combines high-speed with low dissipations both for training and operation.

However, currently only supervised learning of an opto-magnetic network with two synapses was realized and practical realization of opto-magnetic neural networks remains a challenge. In this letter we demonstrate an opto-magnetic neural network that can learn patterns in noisy data and subsequently recognizes these patterns in unseen data. To achieve this challenge, extending the number of synaptic weights was crucial. To this end, we first experimentally demonstrate the realization of 9 optomagnetic synapses in a Co/Pt multilayer structure, which could all be continuously and deterministically controlled, as well as probed, by picosecond laser pulses. Second, we detail an iterative training algorithm for positive synaptic weights, which requires only global feedback. Subsequently, this algorithm is used to train the opto-magnetic synaptic weights, using noisy characters encoded in binary 3x3 images as input. Eventually, we demonstrate successful recognition of unseen characters with the trained optomagnetic neural network. We conclude with a short summary and an outlook to future work.

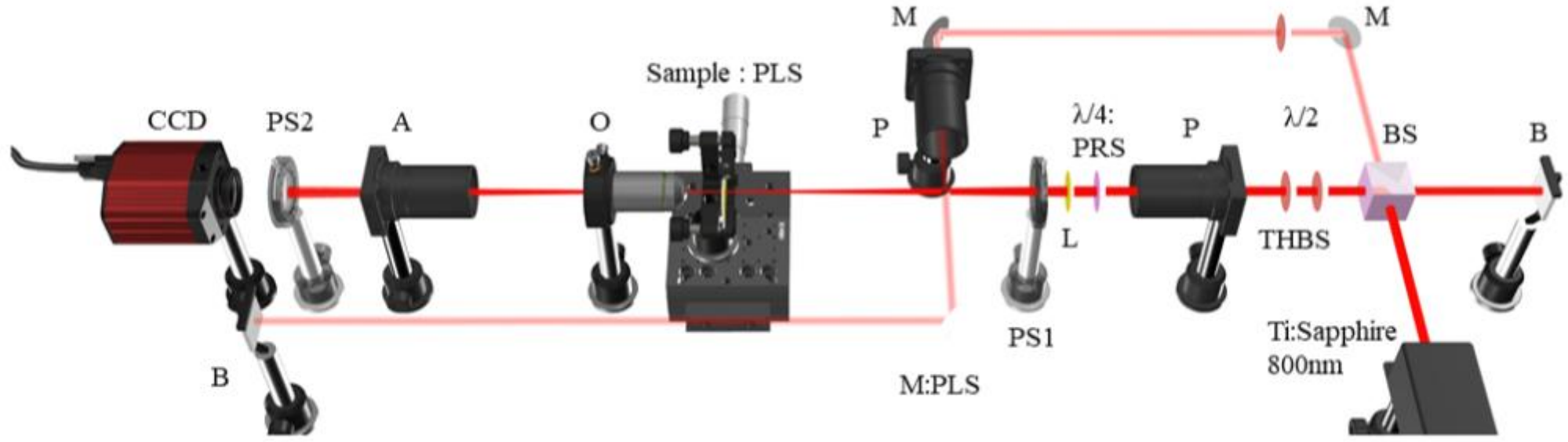

**Figure 1: Schematic illustration of the experimental setup. An optical beam produced from an 800nm Ti: Sapphire laser system is split in a strong pump and weak probe beam by a beam splitter (BS). The pump beam is shaped to a flat-top profile (THBS), circularly polarized (λ/4: PRS) and focused (L) before hitting the sample. A programmable shutter (PS1) releases a finite amount of pump pulses to change the magnetic state of the sample. A second shutter (PS2) prevents the pump beam from reaching the CCD camera. The sample is mounted on a programmable linear stage (PLS) to select different positions. A transferrable mirror (M: PLS) transports the probe beam in and out of the pump path as desired during operation. Using polarizer (P), analyzer (A) and an objective (O), the magnetic domains of the sample are detected on the CCD. The beam stopper (B) is used to park the probe beam when outside the pump path. Half wave plates (λ/2) are used to control the intensity of the beams.**

The experimental setup used to realize the-opto-magnetic neural network is schematically shown in **Figure *1***. We used laser pulses with a wavelength of 800nm, 1kHz repetition rate and 4ps pulse-width from a Ti: Sapphire amplified laser system. An 80:20 beam splitter (BS) is used to create pump and probe Gaussian beams for control and detection of the synaptic weights, respectively. The pump beam shape with a Gaussian waist of 102µm was altered using a TOP-Hat beam shaper (THBS) for better optical control of the magnetic states, i.e. synaptic weights. The probe beam of 4mm diameter was exposed to a monochrome CCD camera for 10ms after it passed through the CoPt sample and a 10× objective lens (O), placed between two nearly crossed polarizers (P and A). The sample used was a multilayered stack, grown on a quartz substrate with a sputter deposition technique. With Glass/Ta(3)/Pt(3)/Co(0.6)/ Pt(3)/MgO(2)/Ta(1) as composition (all bracketed terms are in nano-meter), it has strong out-of-plane anisotropy and a coercivity of about 20mT. A programmable linear stage-based mirror (M: PLS) guides the linearly polarized probe beam to the sample, otherwise, it is used to park the probe beam on a beam stopper (B) screen during the pump exposure. The number of pump pulses on the sample is controlled by a programmable shutter (PS1). When the pump pulses are fired on the sample, another conjugate programmable shutter (PS2) blocks the pump beam before the camera and protects its pixels from the powerful pump pulses. These pulses address nine different weight areas on the sample to build a perceptron network and these areas are accessed by translating the sample using a programmable linear stage (Sample: PLS) in the pump path. Two half-wave plates (λ/2) in pump and probe beam paths, as shown in **Figure *1*** control the beam energies on the sample, yielding approximately 5 mW and 1 mW average power for pump and probe beam, respectively. During the network training, a quarter wave-plate mounted on a programmable rotating stage (λ/4: PRS) allows the pump beam to interchange between right and left circular polarization states for increasing and decreasing the magnetic domains of the sample. The pump beam is focused onto the sample by a lens with a focal length of 40 cm.

To demonstrate continuously adaptable opto-magnetic synapses, we exploit that the laser intensity $I_{out}$ detected in the CCD camera can be written as

$$I_{out} = K\ M\ I_{in} \tag{1}$$

Here $K$ is a material dependent parameter related to the Verdet constant, $M$ the net magnetization of the sample, and $I_{in}$ the intensity of the incident light. Optomagnetic synapses can thus be demonstrated by regarding $I_{out}$ ($I_{in}$) as output (input) and weights $w = KM$, where $M$ is adaptable via optomagnetic effects. Similar as in previous studies[26], we exploit here that Co/Pt thin films exhibit multishot helicity-dependent control of magnetization, allowing to reversibly control its magnetization. Here, this is demonstrated for nine spatially separated positions of the sample. To this end we irradiate the sample with packets of 50 right circularly polarized pump pulses, until a total of 2500 pulses is reached. Subsequently, the helicity of the pump pulses is reversed and again a sequence of 50 pulse packets of 50 pulses each is irridiated onto the sample. This procedure is repeated for nine areas by translating the sample in the pump path. After the exposure of every pulse packet, an image is recorded by bringing the probe beam in to the pump path. Technically, the integrated output intensity per synapse is determined in LabVIEW software by driving two cursors over a section of a continuously streaming sample image on the CCD. Such a sample image typically covers a 166×128 $\mu m^2$ physical area. The cursors sequentially enclose pump written spots, taking snapshots that cover 16.5(W)×15.5(H) $\mu m^2$ areas on the sample. To determine the induced contrast change, a background image is subtracted for convenience. The synaptic weights evaluated in this way from all nine areas, $w_{i=1,\dots,9}$, are plotted as a function of the number of incident pulses in **Figure 2**a. The evolution of the weights shows deviations from ideal linear dependence and is not fully homogenous across the different chosen sample areas (synapses). This may stem from laser fluctuations and/or inhomogenieties of the sample. However, as we will show below, this intrinsic noisiness does not limit the performance of the network and its ability to realize pattern recognition.

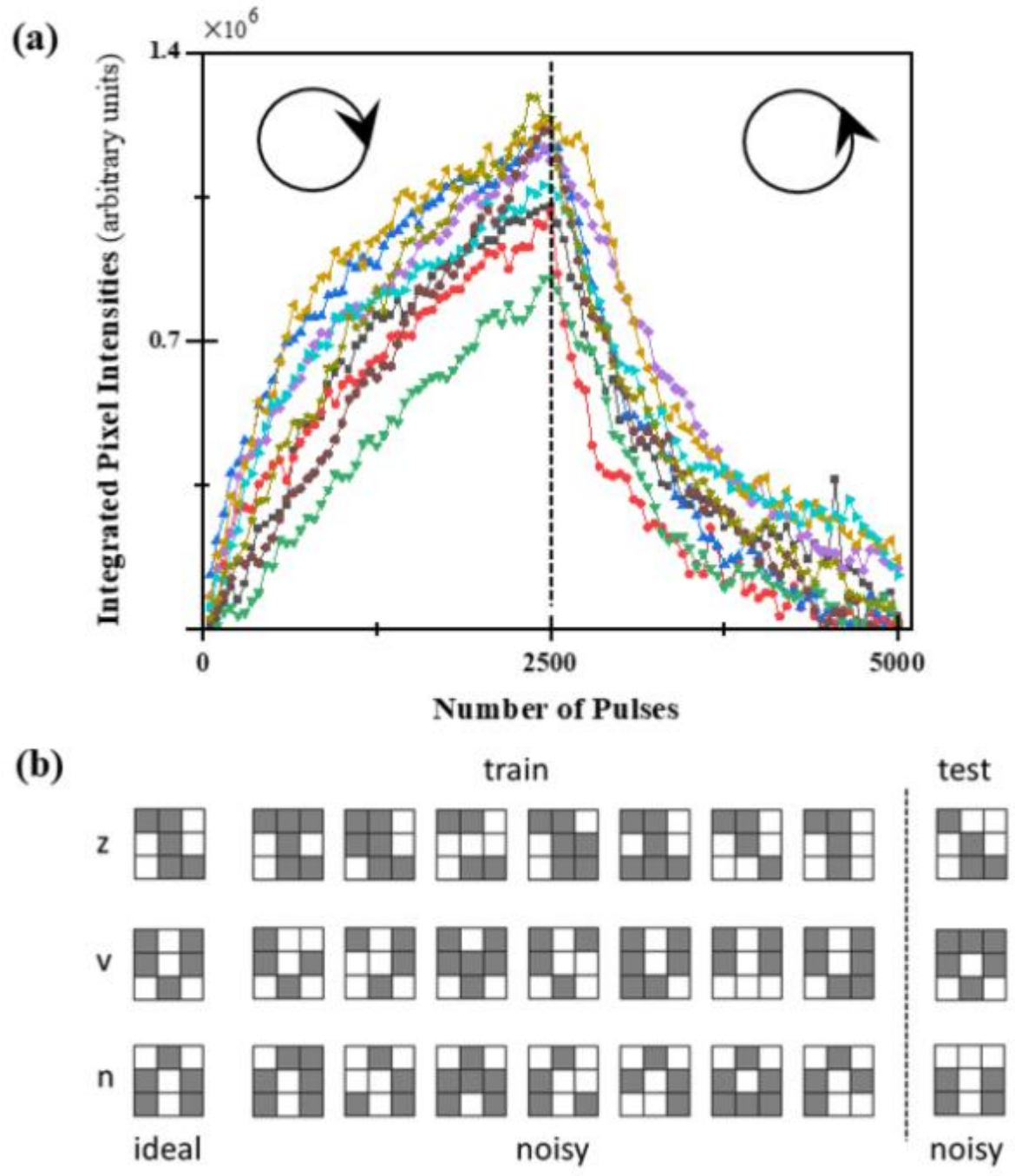

**Figure 2: (a) Changes of magneto-optical contrast as a function of the number of circularly polarized pump pulses. After 2500 pulses the helicity is changed from left to right circularly polarized light. Different colors correspond to nine different areas in the sample, corresponding to nine opto-magnetic synapses. In all cases reversible control of magneto-optical contrast is observed. (b) illustration of the binary 3x3 patterns in the data set, adapted from[3]. 24 patterns (left) are used for training and 3 (right) for recognition.**

To demonstrate training and pattern recognition with a network of nine opto-magnetic synapses we consider a dataset shown in **Figure 2**b, comprising 3x3 binary images representing the characters 'z', 'v' and 'n', see also[3]. Eight noisy versions are created by flipping one bit of the ideal pattern, creating a total data set of 27 images. This data set is linearly separable, which can be seen by considering the Hamming distance between patterns, which is maximally 2 for patterns in the same class and minimally 4 for patterns in different classes. For each of the classes, one noisy pattern is left out as test set, the remaining 24 patterns are used for training.

The input of the network comprises probing light on each synapse: $x_i^\mu$ =0/1 = light/no light, for input $i = 1, \dots, 9$ and pattern $\mu = 1, \dots, 27$. The output is the integrated intensity detected for each synapse, summed over all synapses $O^\mu = \sum y_i^\mu = \sum w_i x_i^\mu$. In the experiment the 3x3 images are encoded in a 1D array. Distinct from our previous study[26], where the input condition $x_i^\mu$ =1/0 was realized by (not) blocking the probing beam with a physical shutter, in the current setup the input probing light is always present and the output $y_i^\mu$ is simply excluded when $x_i^\mu = 0$ in the LabVIEW controlling and detection software of the setup.

Since $O^\mu$ in the experiment is strictly positive, we train the network with a positive perceptron rule previously derived for optical perceptrons[27]. For two classes $C_1$and $C_2$ and threshold $b$, the weight updates are determined by

$$\Delta w_i = \eta \begin{cases} 0 & O^\mu > b \quad \mu \in C_1 \\ 0 & O^\mu \leq b \quad \mu \in C_2 \\ x_i^\mu & O^\mu \leq b \quad \mu \in C_1 \\ -x_i^\mu & O^\mu > b \quad \mu \in C_2 \end{cases}, \qquad (2)$$

where $\eta$ is the learning rate. For actual training we exploit the fact that the dataset is linearly separable, such that weights can be trained for the patterns 'z', 'v' and 'n' sequentially. For a given class of patterns, training is done until the desired output $O_*^\mu > b$ holds for $\mu \in C_1$ and $O_*^\mu \leq b$ for $\mu \notin C_1$. For example, if we train the network to characterize 'v', $C_1$ comprises all (noisy) patterns of 'v', while $C_2$ comprises all patterns 'n' and 'z' that are not in $C_1$. The training proceeds in an iterative procedure (sometimes called online training) where weight changes are considered after each pattern to reduce external storage demands. Weights are changed when the output is undesirably below or above threshold, otherwise they remain unchanged and the next pattern is presented to the network. In the experiment $\eta$ depends on the number of pump pulses per learning step, which is controlled by the electronic shutter that exhibits opening times varying between 15 and 25ms; $\mathrm{sign}(\eta)$ is determined by the quarter wave plate (see **Figure *1***).

Experimental results of the pump-induced changes of the synaptic weights during training are shown in **Figure *3***, where patterns in the class 'v' are learned. The bottom image shows the bright state (fully polarized or 'up'). Prior training, each optomagnetic synapse is irradiated with 2500 right circularly polarized pump pulses to create the initial dark (or 'down') state for each synapse. Subsequently, the algorithm feeds patterns causing changes of the weights in accordance with Eq. (2). After about 400 training steps training is completed.

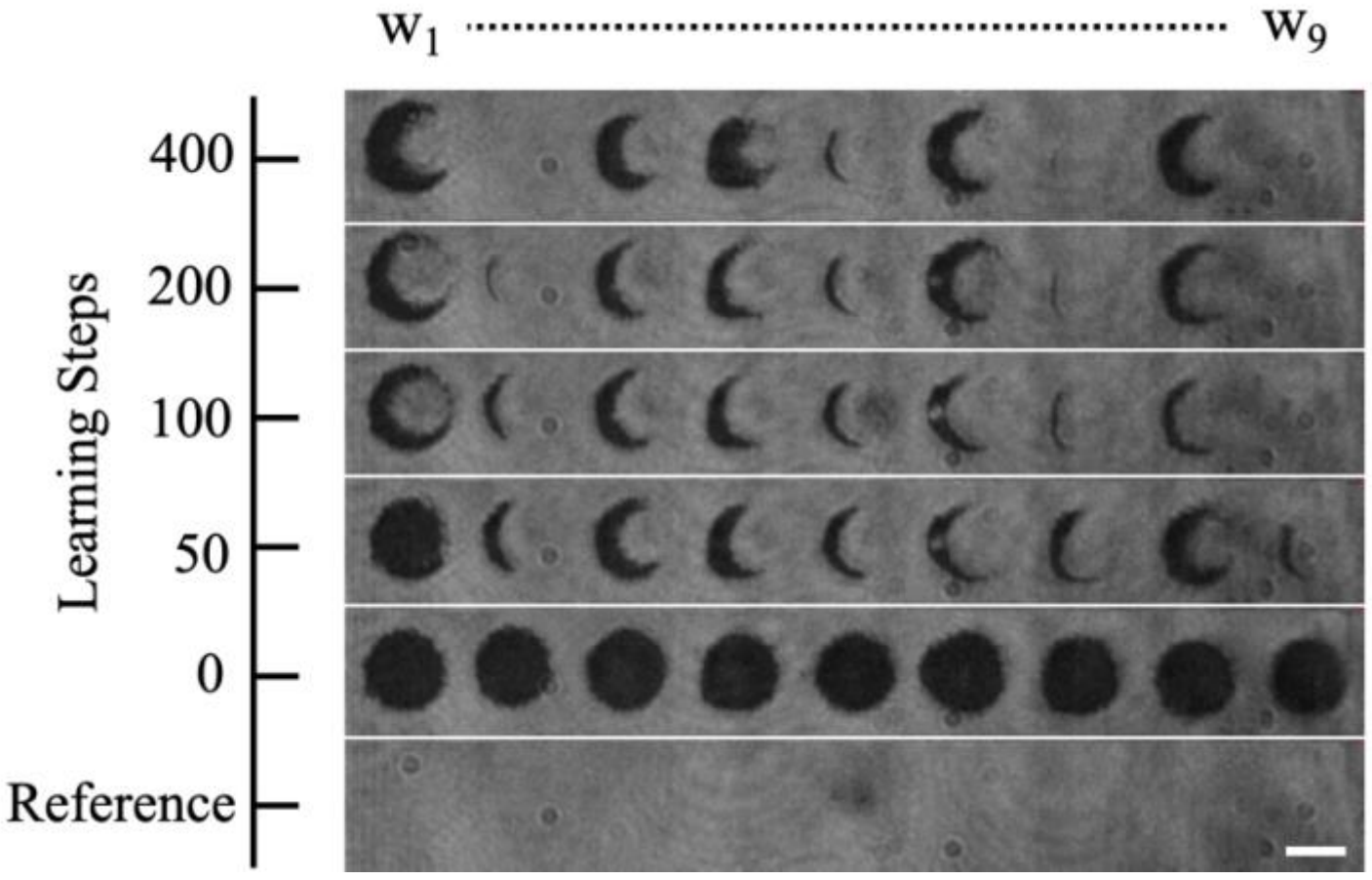

**Figure 3: Changes of the optomagnetic synapses during training. The initial state is defined by saturating the synapses from the fully polarized background with 2500 right circularly polarized pulses. Scale bare: 10 µm.**

In order to follow the classification error of the network on the training data, **Figure *4*Figure *1***a shows the integrated output $O^{\mu}$ as a height bar plot as a function of the training step. To assist analysis, the bars are colored blue (red) when the classification is correct (wrong). The horizontal black-solid line indicates the threshold $b$. The sequence of learning steps is such that first the 8 training patterns 'z' are provided, subsequently the 'v' and then the 'n' training patterns. The sequence of these 24 training patterns is repeated until for all training patterns the desirable output is obtained. **Figure *4***b shows the results on the first sequence of patterns, where for all three types of patterns erroneously classified patterns are found. **Figure *4***c shows the results after training, where all patterns 'v' exhibit $O^{\mu} > b$ and all other patterns have $O^{\mu} < b$. Hence, the network correctly recognizes all the patterns of the training set.

As last step of the experiment, the test data is presented to the network. The result is shown in **Figure *4***d by dark blue textured bars, together with all data from the training set. It is observed that the network also recognizes the 3 test patterns successfully. Hence, we have experimentally demonstrated both training and pattern recognition with an opto-magnetic neural network.

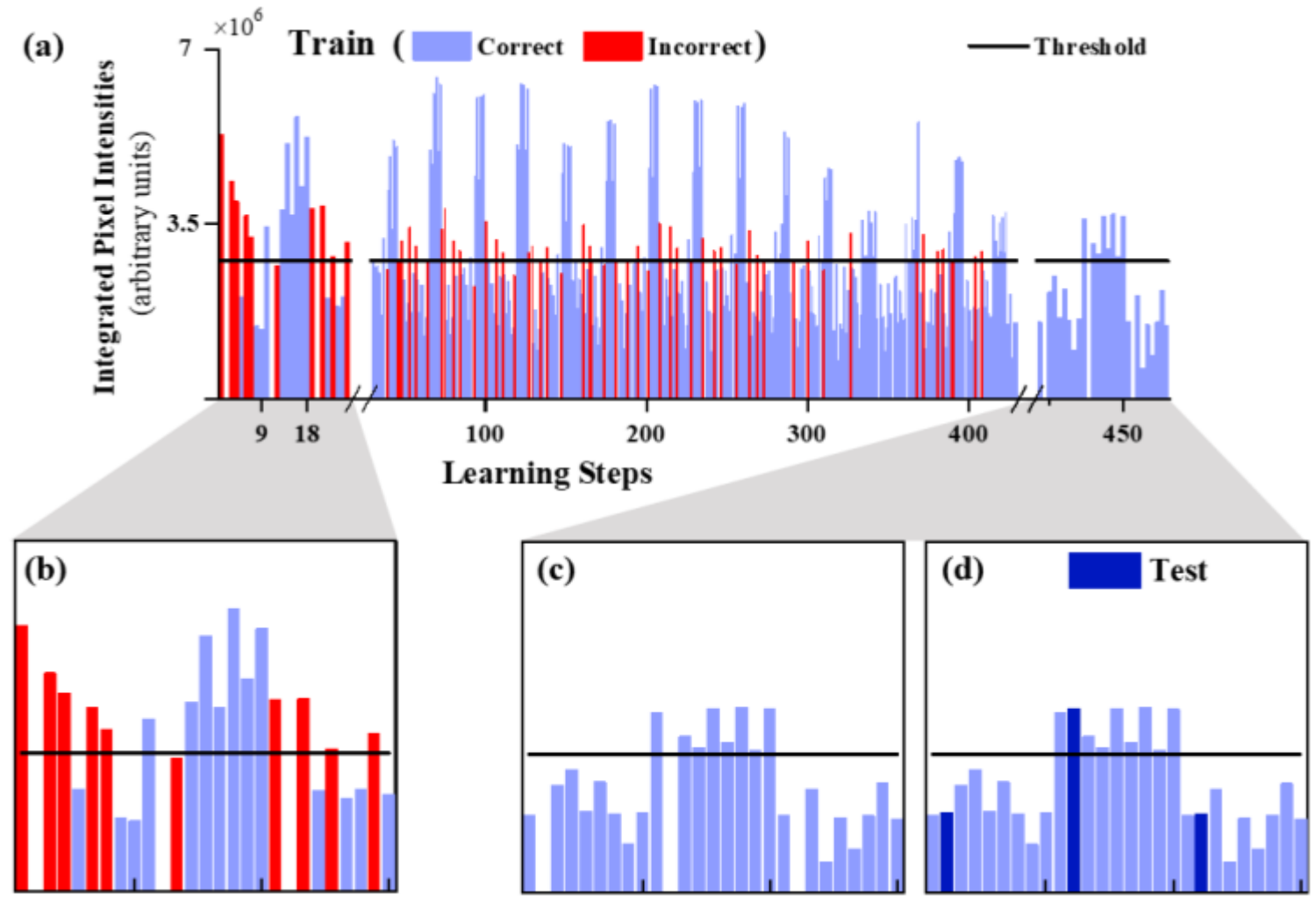

**Figure 4: (a) Height bar plot of integrated output of the network as a function of training step and during operation. Blue and red bars indicate correctly (erroneously) recognized patterns. The horizontal line indicates the threshold. (b) Initially, many patterns are not recognized appropriately. (c) after about 400 training steps, all patterns are correctly above and below threshold. (d) the test patterns, which were not part of the training are also recognized correctly (dark blue textured bars).**

Next, we estimate the energy efficiency of the demonstrated opto-magnetic neural network. We estimate the absorbed fluence from the largest diameter of the laser-written spots, which is 12.5 µm in **Figure *3***. By considering the power, repetition rate and Gaussian width, the laser fluence is estimated as 63 mJ/cm$^2$, the absorbed energy per spot then becomes 31 pJ/synapse/pulse. The writing energy can be further reduced by considering tighter focusing and smaller spots, yielding favorable scaling down to fJ for nanoscale synapses. Moreover, recent experiments show that in nanomagnets the ultrafast switching intrinsically operates at reduced laser fluence[28].

In summary, opto-magnetic neural networks offer an appealing approach to combine high data transfer rates with low energy dissipation both for training and operation. An advantage of optical processing is that it can be operated massively parallel, though in our current experimental setup the synapses are accessed sequentially. Future work might therefore focus on parallel training and operation, which can be implemented by exploiting microlens arrays, which splits the main beam in many focal spots[29], while parallel and selective weight adaption might be done using the principles of wavefront shaping. To approach the data rates accessible in photonics, it is desirable to develop integrated photo-magnetic circuits, benefiting from recent developments with optical MRAM[30,31] and eventually also novel solid-state laser technology[32].

We would like to acknowledge S. Horst and M. Veis for stimulating discussions, K. T. Yamada for supplying the sample, and F. Ando and T. Ono for sample growth. This work received funding from the European Research Council: ERC grant agreement no.856538 (3D-MAGiC), the Nederlandse

Organisatie voor Wetenschappelijk Onderzoek (NWO), and the Nationale Wetenschaps Agenda (NWA, Startimpuls GreenICT). J.H.M. acknowledges funding from the Shell-NWO/FOM-initiative "Computational sciences for energy research" of Shell and Chemical Sciences, Earth and Life Sciences, Physical Sciences, FOM and STW.  Technical support from Tonnie Toonen and Chris Berkhout is gratefully acknowledged. The experimental setup in Figure 1 was drawn with an online tool at 3DOptix.

Data availability statement

The data that support the findings of this study are available from the corresponding author upon reasonable request.